\documentclass[slac_one]{revtex4}
\usepackage{graphicx}
\usepackage{fancyhdr}
\begin{document}

\title{{\small{2005 International Linear Collider Workshop - Stanford,
U.S.A.}}\\ 
\vspace{12pt}
Fast Detector Simulation Using Lelaps,
Detector Descriptions in GODL} 

%

\author{Willy Langeveld}
\affiliation{SLAC, Stanford, CA 94025, USA}
%

\begin{abstract}
Lelaps is a fast detector simulation program which reads StdHep generator files
and produces SIO or LCIO output files. It swims particles through detectors taking into
account magnetic fields, multiple scattering and dE/dx energy loss. It simulates
parameterized showers in EM and hadronic calorimeters and supports gamma conversions
and decays. In addition to three built-in detector configurations, detector descriptions
can also be read from files in the new GODL file format. 
\end{abstract}

\maketitle

\thispagestyle{fancy}


\section{INTRODUCTION} 
Lelaps is a fast detector simulation that swims particles through
detectors with magnetic fields, accounting for multiple scattering
and energy loss. It produces parameterized showers in EM and hadronic
calorimeters, supports decays of certain short-lived particles (``V'' decays)
and converts gammas. Lelaps performance is on the order of 1 typical
(e.g. ZZ) event/second at 1 GHz, with everything turned on and writing
an LCIO output file.

Lelaps consists of a set of C++ class libraries and a main program,
which itself is called lelaps. The main class library is called CEPack,
the actual simulation toolkit. It reads StdHep generator files and produces
SIO or LCIO output files. Lelaps has built-in support for LDMar01,
SDJan03 and SDMar04. It also reads detector geometries in the new GODL
format.

For details on Lelaps, see references~\cite{LCWS04} and~\cite{freehep}. In
the next section, a brief reminder is provided. The rest of this paper
gives an introduction to the GODL detector description language.

\section{A LELAPS REMINDER}

\begin{figure}
     \begin{center}
     \vspace*{.2cm}
     \begin{tabular}{cc}
     \mbox{\includegraphics[width=6cm]{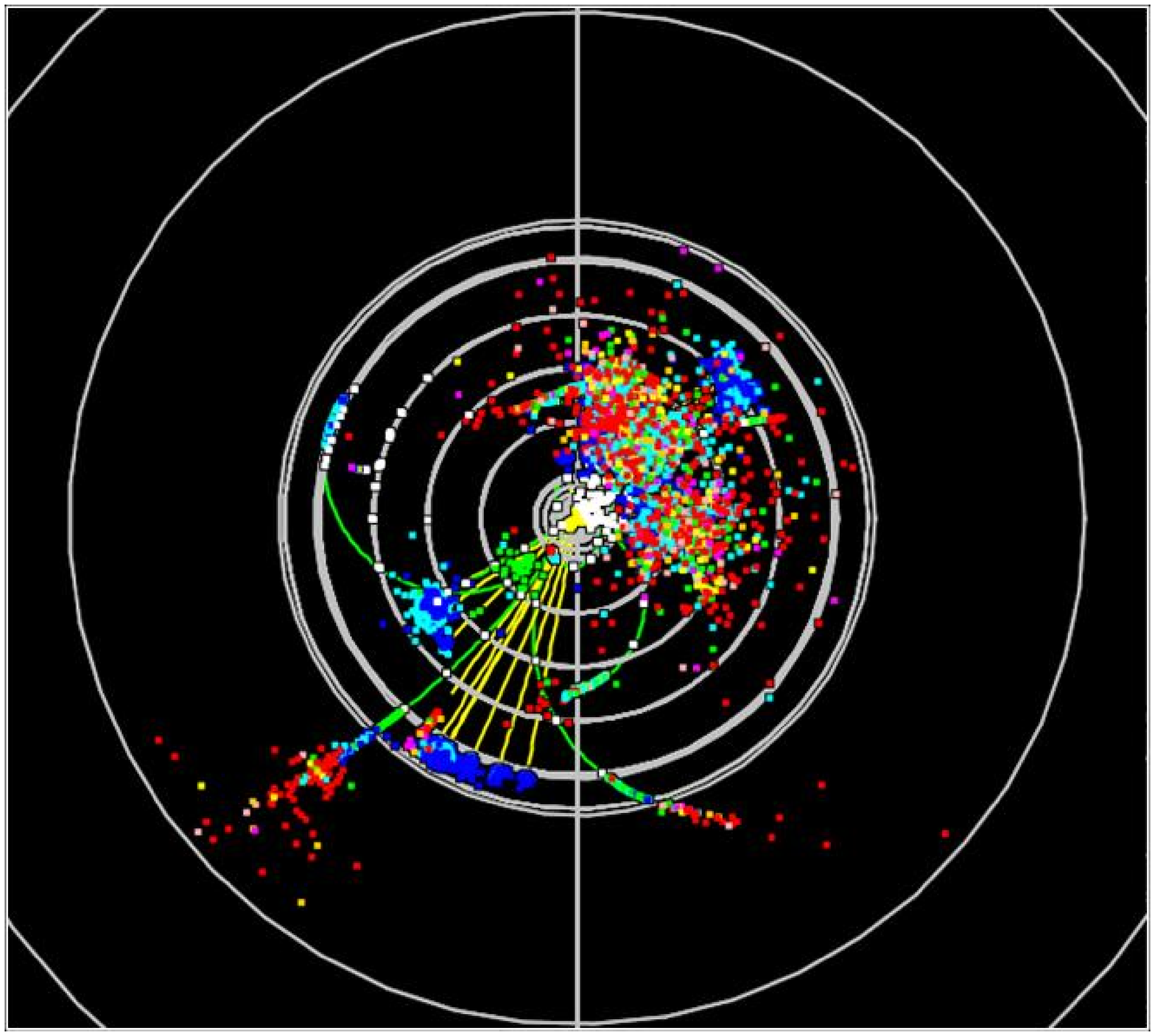}}
     \mbox{\includegraphics[width=6cm]{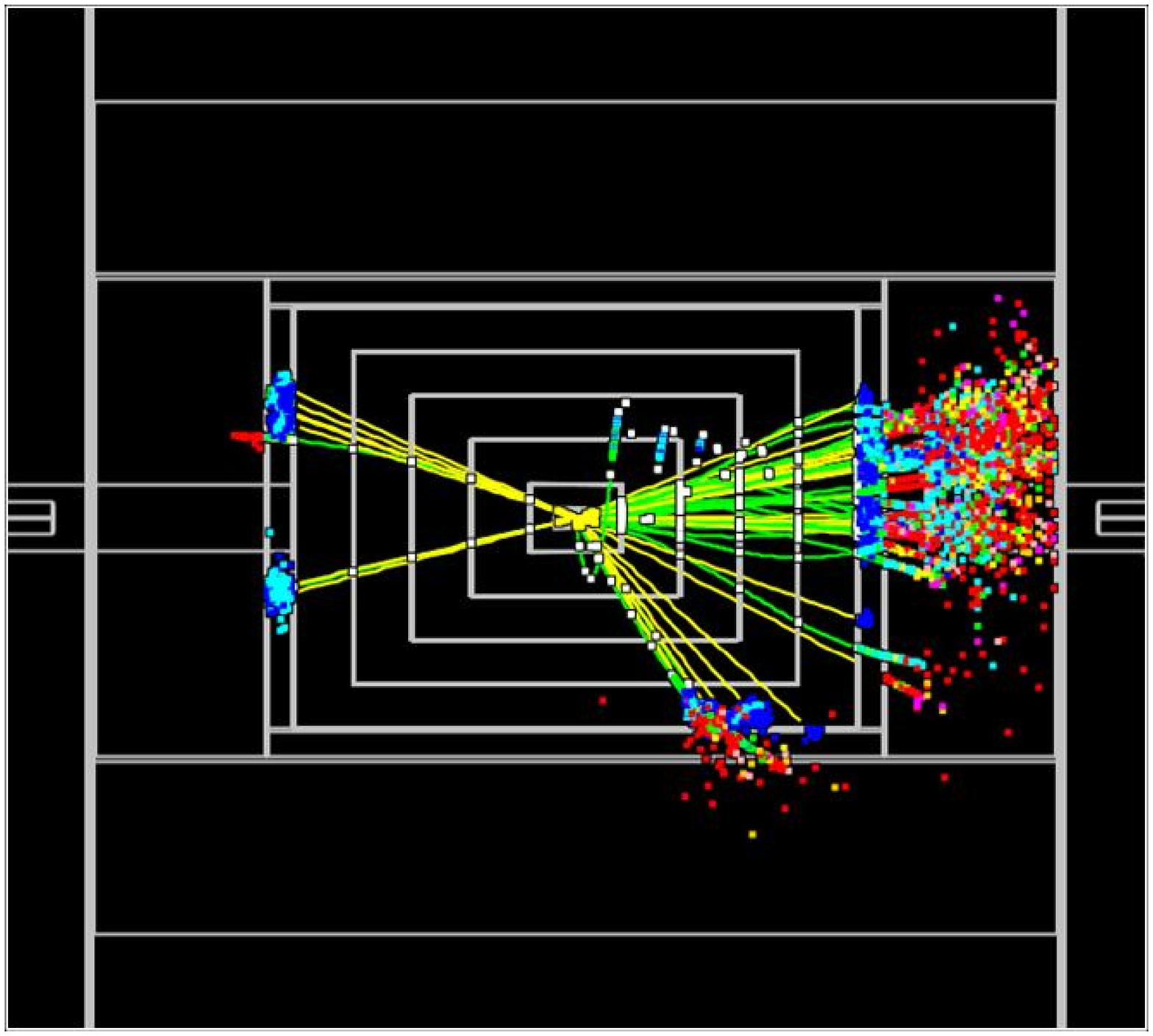}}
     \end{tabular}
     \end{center}
     \caption{ZZ event in the SiD detector, as simulated by Lelaps.} 
     \label{fig:sidid}
     \end{figure}

\begin{figure}
     \begin{center}
     \vspace*{.2cm}
     \begin{tabular}{cc}
     \mbox{\includegraphics[width=6cm]{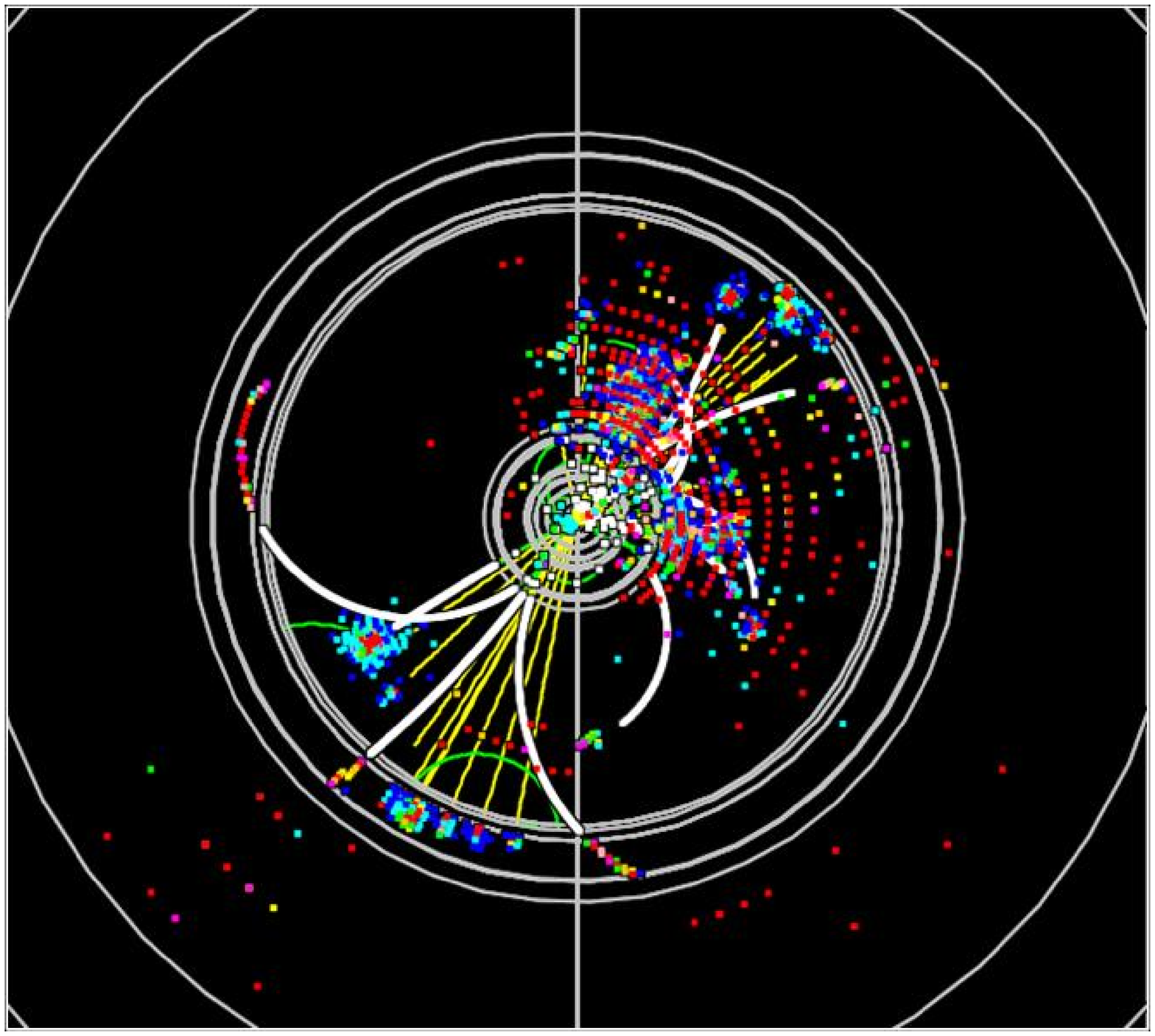}}
     \mbox{\includegraphics[width=6cm]{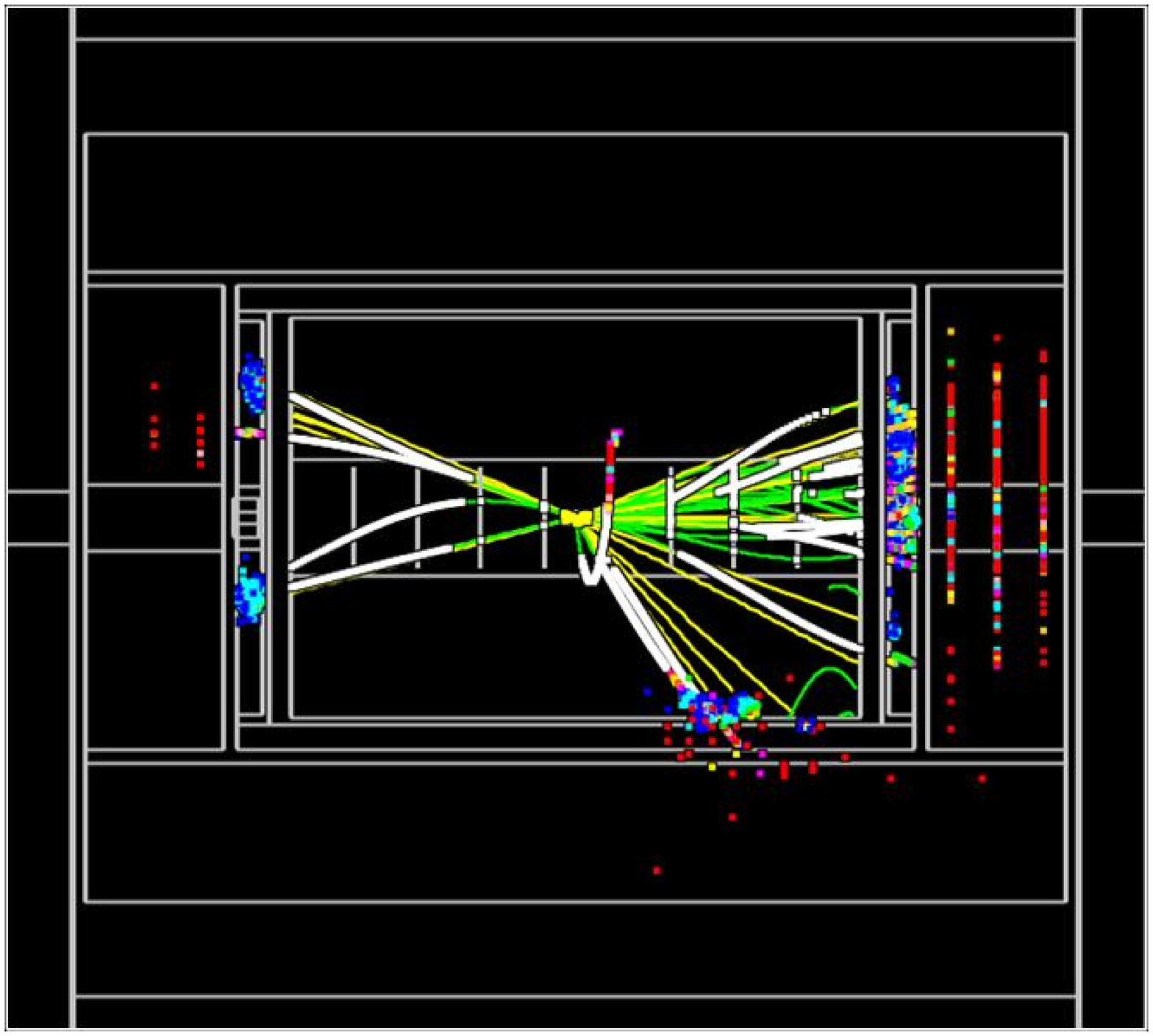}}
     \end{tabular}
     \end{center}
     \caption{ZZ event in the LD detector, as simulated by Lelaps.} 
     \label{fig:ldid}
     \end{figure}

The main class library for Lelaps is called CEPack and it contains the simulation
tool kit. It deals with geometry, materials and tracking particles through the geometry
while taking into account magnetic fields, multiple scattering, energy loss, parameterized
showers in EM and hadronic calorimeters, particle decays and gamma conversions.
A Lelaps-simulated ZZ event in the SiD detector is shown in figure~\ref{fig:sidid}. The
same event simulated in the LD detector is shown in figure~\ref{fig:ldid}.


\subsection{Materials}
Specifying materials is very easy in Lelaps. All elements are built in with 
default pressure and temperature for gasses or density for solids and liquids.
Any compound can be specified by chemical formula and density or (for gasses)
temperature and pressure. Mixtures can be created by mixing elements and compounds
by volume or by weight. All needed quantities are calculated automatically. This
includes constants needed for multiple scattering and energy loss, radiation lengths,
interaction lengths and constants needed for shower parameterization.

The Lelaps distribution comes with a little program called matprop that
allows one to view various material properties. An online version of
matprop is available~\cite{matprop}.

\subsection{Multiple Scattering and Energy Loss}
Tracking is performed by taking steps along a linear trajectory with endpoints
on a helix, such that the sagitta stays below a certain maximum.
After each step, the amount of material traversed is checked: if enough material
was traversed, multiple scattering and energy loss is performed and track parameters
are updated. When an intersection occurs within a step, the fractional step is executed,
the volume is entered, and the remaining fraction of the step follows.

Multiple scattering is performed using the algorithm of Lynch and Dahl~\cite{dahl}.
Material is saved up along the track until there is enough. dE/dx is calculated
using the methods by Sternheimer and Peierls~\cite{sternheimer}. All constants are
precalculated by the material classes.

\subsection{Shower Parameterization}
     \begin{figure}
     \begin{center}
     \includegraphics[width=8cm]{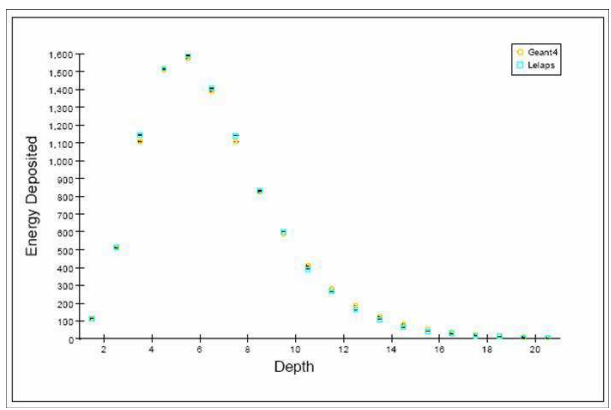}
     \end{center}
     \caption{Comparison of longitudinal shower distribution as simulated by Geant4 and Lelaps.} 
     \label{fig:compG4id}
     \end{figure}
Electromagnetic showers are parameterized using the algorithms of Grindhammer and
Peters~\cite{grindhammer}.
Calorimeters are treated as homogeneous media. The longitudinal shower profile is given
by a gamma distribution with coefficients depending on the material (Z) and energy.
The profiles are fluctuated and correlations between the coefficients are taken into account.

For each step of one radiation length, a radial profile is computed consisting of
two distributions, one describing the core of the shower and the other the tail.
The energy to be deposited is divided into spots thrown in radius according to the
radial profile, and uniformly in $\phi$. Roughly, about 400 spots are generated
per GeV of shower energy and reported as hits.

Hadronic showers are parameterized in a similar way, with some modifications.
The location where the shower starts is simulated using an exponential law with
attenuation given by the interaction length. The longitudinal profile uses the Bock
parameterization~\cite{bock}. A combination of two gamma distributions, one using radiation
lengths and the other interaction lengths, is used. The Bock parameterization does
not specify radial profiles. For the moment we use a radial profile similar to
Grindhammer and Peters (for EM showers) but with radiation lengths replaced by
interaction lengths and faster spread with depth. The parameters still need to
be fine-tuned.

These parameterizations were compared to results from Geant4~\cite{geant4}.
In general pretty good agreement was found for EM showers (see figure~\ref{fig:compG4id}).
Hadronic showers agree pretty well longitudinally, but not as well radially. Hadronic
shower parameterization has been tweaked since then, but needs further work.

\subsection{Decays and Gamma Conversions}
     \begin{figure}
     \begin{center}
     \vspace*{.2cm}
     \begin{tabular}{cc}
     \mbox{\includegraphics[width=6cm]{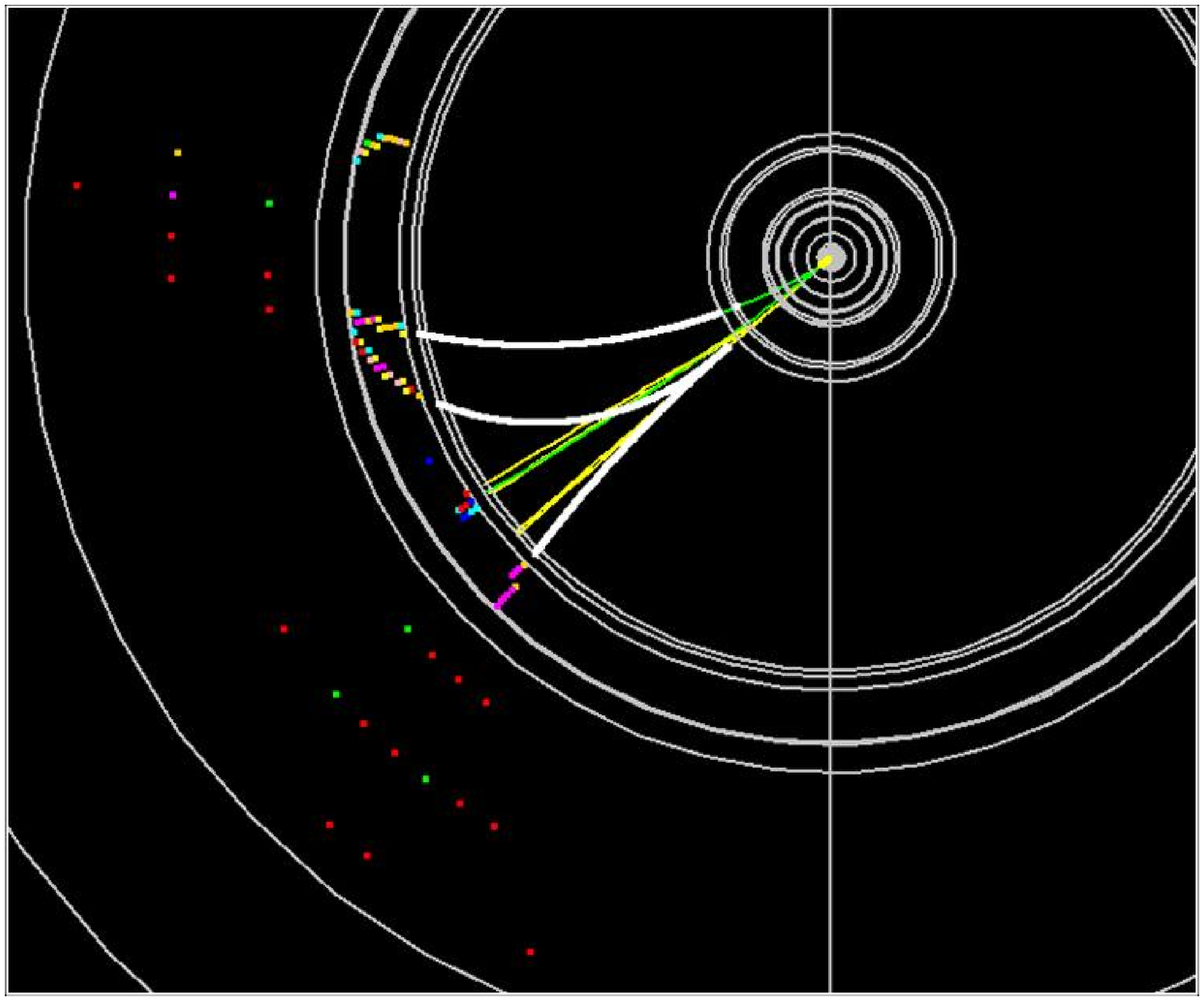}}
     \mbox{\includegraphics[width=6cm]{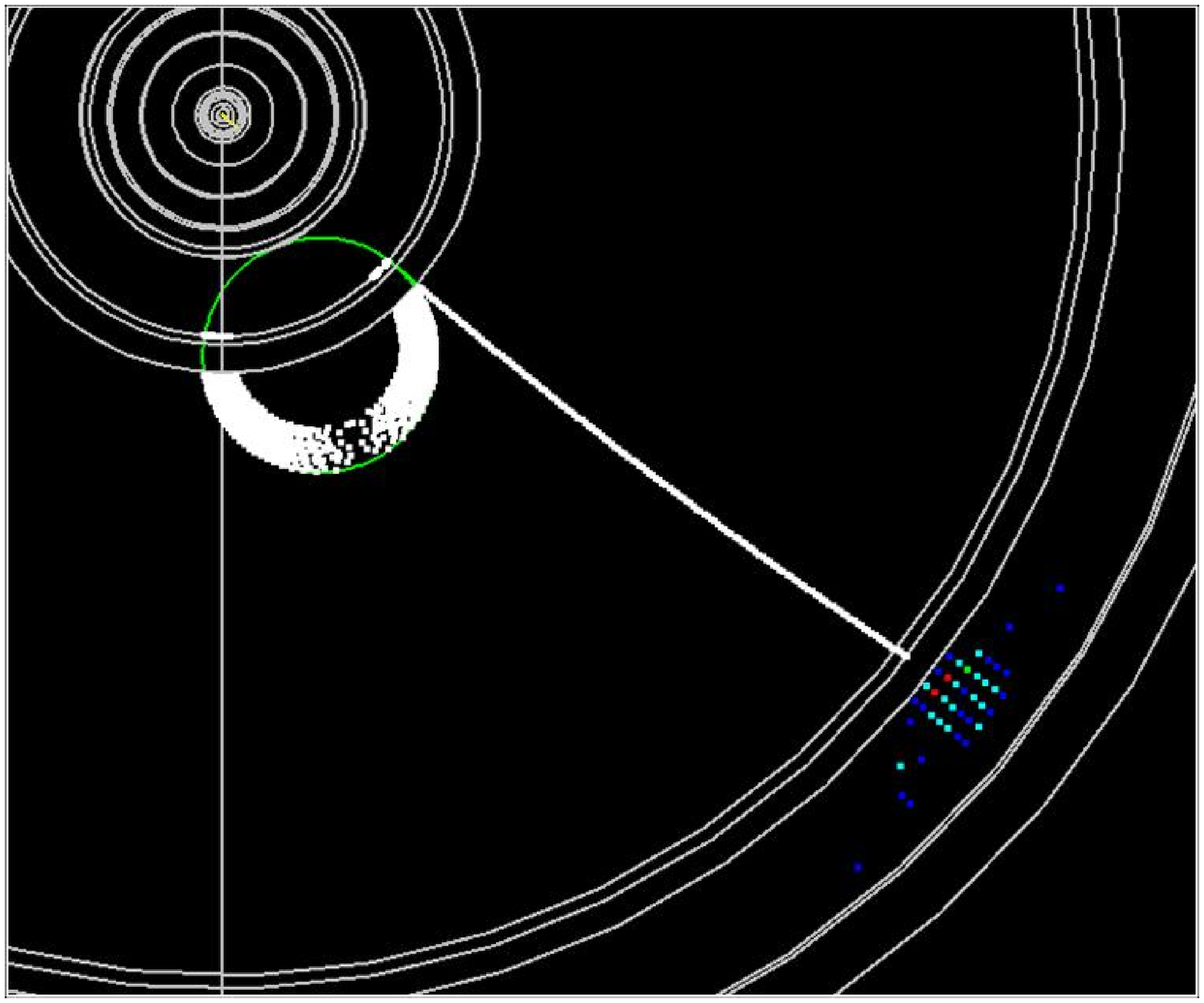}}
     \end{tabular}
     \end{center}
     \caption{Pictures of an $\Omega^-$ decay and a gamma conversion as simulated by Lelaps.} 
     \label{fig:decaygamid}
     \end{figure}
CEPack supports decays of unstable particles and gamma conversions. Supported unstable
particles are $\pi^0$, $K^0_s$, $\Lambda$, $\Sigma^{+/-/0}$,
$\Xi^{-,0}$ and $\Omega^-$. Only decay modes with branching fractions greater than 2\%
are supported (mostly ``V decays''). See figure~\ref{fig:decaygamid}.

\section{GENERALIZED OBJECT DESCRIPTION LANGUAGE (GODL)}
A new feature in Lelaps is the ability to read in detector descriptions written in a new
language called GODL. This language was especially designed to describe all aspects of
detector geometry, materials, detector-subdetector relationships, as well as the output
formats for any hits produced in sensitive parts of the detectors. We will start with a
subsection on the features of GODL and conclude with the current status of implementation.

\subsection{Features}
GODL is a more or less complete programming language with variables and operations. This
means that one does not have to hard-code dimensions and locations but can instead
compute dimensions and locations from previously defined ones. GODL has control constructs
(loops, if) which make repetitive operations easier. There are built-in math functions,
which allows calculating derived quantities (e.g. the tangent of an angle).
The language is human readable and editable (like XML but more so) and portable (since it
is in plain text). The interpreter catches mistakes (like XML but better).

GODL knows about units and enforces them. For example, one can mix microns and meters.
One can define new units based on built-in ones and use them. The interpreter enforces
consistent usage of units in operations and function calls.

GODL has list objects. Built-in objects describe materials, objects, placements etc.
GODL supports volume hierarchies. This saves simulation time by allowing embedding
of sub-detector elements into mother volumes.
GODL allows specification of arbitrary calorimeter segmentation and encoding of tracker and
calorimeter ID's. For this, it has a built-in, PostScript-like, parser which ``compiles'' a
suitable ID code specification to byte-code for fast execution. This allows one to change
encoding and/or segmentation without modifying the simulator source code.

GODL supports levels of detail. This allows one to use a low level of detail
for fast simulation and higher level of detail for full simulation, all encoded in the same
GODL source file.

GODL comes with a simple API which currently comprises 11 virtual methods, which are all
fully implemented in Lelaps. Lelaps comes with a GODL-to-HepRep converter (these HepReps
require the newest Wired4 for viewing). A Geant4 implementation is planned.

GODL supports (in principle) all Geant4 solid types (with the possible exception of BREPS).
Not all of them are implemented at this time, but it would be easy to do.
And finally, GODL supports (in principle) combinatorial (``boolean'') geometry
(``CSG'' in Geant4). This is not yet implemented, but should be straightforward.

\subsection{Variables and Arrays}
GODL is an extensible typeless programming language. Type is determined by assignment:
\begin{verbatim}
   a = 2.4;      # real
   b = 2;        # integer
   c = "text";   # string
   d = true;     # boolean
\end{verbatim}

It has variables and operations that can be performed on them:
\begin{verbatim}
   a += 1;
   b = a * 12;
   d = c + " more text";
   e = false;
   b = e != true;
\end{verbatim}

It also supports array-like constructs:
\begin{verbatim}
   i = 5; foo.i = 12;       #  Same as foo.5 = 12;
\end{verbatim}
These arrays are much like ``stem'' variables in REXX.

\subsection{Operators}
GODL supports the following set of operators, although some cannot be used in some contexts:
\begin{verbatim}
   + - * / = += -= *= /= == < > <= >= != ! && ||
\end{verbatim}
The meaning of these operators is just like in C. Note the absence of $++$ and $--$ operators.

There is also a reference operator $@$:
\begin{verbatim}
   a = 12;
   b = @a; print(b, "\n");
\end{verbatim}
which would print out:
\begin{verbatim}
   @a->(12)
\end{verbatim}
This is useful for referencing objects multiple times without recreating them, see later.

\subsection{Built-in Functions}
GODL knows about the usual set of math functions:
\begin{verbatim}
   exp, log, sqrt, pow, cos, sin, tan, acos, asin, atan, atan2, cosh, sinh, tanh,
   acosh, asinh, atanh, log10, abs, fabs, ceil, floor and mod.
\end{verbatim}

In addition there is a list function:
\begin{verbatim}
   a = list(a, b, c, d);    # Creates unnamed list
\end{verbatim}
and a print function:
\begin{verbatim}
   print(a, "\n");
   print(a, "\n", b, "\n");
\end{verbatim}

It is possible to use a GODL parser as a ``shell''. When arguments are provided, argc and
argv work more or less like they do in C.

There is also a ``unit'' function, see later.

\subsection{Control Constructs}
GODL has a limited set of C-style control constructs:
\begin{verbatim}
   for (i = 0; i < 25; i += 1) {   ...   }

   while (true) {
      ...
      if (something) break;
   }

   if (a < b) {   ...   }
\end{verbatim}

\subsection{List Objects}
Variables can be list objects. To construct a generic list with name ``foo'':
\begin{verbatim}
   a = foo(a, b, c, d);
\end{verbatim}
Lists can contain objects of any type, including other lists. To add objects to a list:
\begin{verbatim}
   a += e;
   a += f;
\end{verbatim}
Note that this is not necessarily the same as:
\begin{verbatim}
   a += e + f;
\end{verbatim}
which would first add f to e and then the result to a. If e and f are list objects,
this adds to ``a'' a single list ``e'' which in turn contains ``f''.

Note that the GODL parser built into Lelaps will disallow any named lists
whose names do not match one of the set of lists it recognizes (see later).
 
\subsection{Units}
Variables can have units, and units are enforced across operations and in arguments
to function calls and list objects:
\begin{verbatim}
   m = _meter;     # _meter is a built-in unit
   unit("m");      # Declare as unit
   a = 2 m;
   b = 12 * a;
   area = a * b;
   area += 5;      # Error: incorrect units
   d = cos(area);  # Error: cos() only takes angles
\end{verbatim}
Available basic units are (like CLHEP):
\begin{verbatim}
   _meter, _second, _joule, _coulomb, _kelvin, _mole, _candela, _radian, and _steradian.
\end{verbatim}
Built-in units derived from these are: 
\begin{verbatim}
   _angstrom, _parsec, _barn, _degree, _hertz, _becquerel, _curie, _electronvolt,
   _gram,  _watt, _newton, _pascal, _bar, _atmosphere, _ampere, _volt, _ohm,
   _farad, _weber,  _tesla, _gauss, _henry, _gray, _lumen, and _lux.
\end{verbatim}
One can create new units:
\begin{verbatim}
   m = _meter; g = _gram; # For convenience
   unit("m", "g")         # Declare as units
   gcc = g/cm3;           # New unit of density
   unit("gcc");           # Declare
\end{verbatim}
SI prefixes and powers are automatically converted:
\begin{verbatim}
   a = 1 cm2;             # = 0.0001 _meter squared
\end{verbatim}
There are some built-in constants:
\begin{verbatim}
   _pi (3.14...) (has units of rad)
   _e_SI         (electron charge, 1.6...10-19 C)
   _e (2.71...)  (dimensionless)
\end{verbatim}

\subsection{Miscellaneous Functions and Variables}
For debugging, there are two functions:
\begin{verbatim}
   verbose       prints a lot of debugging information to stdout
   __printvars   prints a list of all variables to stdout
\end{verbatim}
Further, there are some control variables for the print() function:
\begin{verbatim}
   printlevel_: (default 1) controls how much information to print (mostly for for object lists).
   precision_:  controls how many digits are displayed for floating point numbers.
   fieldwidth_: controls how much space a printed number takes up.
\end{verbatim}

\subsection{Materials}
Materials are declared using the element, material or mixture list objects
(note the use of the $@$ operator to pass by reference):
\begin{verbatim}
   Si     = element("Si");
   vacuum = material("vacuum");
   O2     = material(formula("O2"), pressure(1.0 atm), temperature(293.15 K));
   Tyvek  = material(name("Tyvek"), formula("CH2CH2"), density(0.935 g/cm3));
   Air    = mixture(part(@O2, 20.946), part(@N2, 78.084), part(@Ar, 0.934), by("volume"));
\end{verbatim}

\subsection{Volumes and Placements}
In order to construct a geometry, we first define a World Volume:
\begin{verbatim}
   World = cylinder(radius(700.0 cm), length(14.0 m), @vacuum);
\end{verbatim}
To define another volume we use, for example:
\begin{verbatim}
   em_ec_irad  = 21.0 cm;                  em_ec_orad   = 125.0 cm;
   em_b_irad   = em_ec_orad + 2.0 cm;      em_thickness = 15 cm;
   em_b_orad   = em_b_irad + em_thickness; em_nlayers   = 30;
   em_sampfrac = 0.02664;                  em_b_length  = 368.0 cm;

   EM_Barrel = cylinder(name("EM Barrel"), innerRadius(em_b_irad), outerRadius(em_b_orad),
                        length(em_b_length), @SiW, type("emcal"), nLayers(em_nlayers),
                        samplingFraction(em_sampfrac));
\end{verbatim}
Now we add this to World using a placement:
\begin{verbatim}
   World += placement(@EM_Barrel);
\end{verbatim}

We can use loops to do repetitive tasks and if statements for conditionals:
\begin{verbatim}
   Vertex_Barrel = cylinder(name("Vertex Barrel"), innerRadius(v_irad),
                            outerRadius(v_orad), length(v_lenmax));
   for (i = 1; i <= v_nlayers; i += 1) {
      vlen = v_leninner;
      if (i > 1) vlen = v_lenmax;
      Vertex_Barrel.i = cylinder(name("Vertex Barrel " + i), ... );
      Vertex_Barrel += placement(@Vertex_Barrel.i); # Notice hierarchy
   }
   World += placement(@Vertex_Barrel);
\end{verbatim}

\begin{table}[t]
\begin{center}
\caption{Level of detail specification syntax}
\begin{tabular}{|l|c|l|}
\hline \textbf{Level syntax:} & \textbf{Create object when:} & \textbf{Used for:} \\
\hline $<$not specified$>$    & always                  & Fundamental objects that are always present\\
\hline level(min(2))          & level $\geq$ 2          & Detailed objects that should not be simulated\\
                              &                         & at lower levels \\
\hline level(max(4))          & level $\leq$ 4          & Fundamental objects that are replaced with other\\
                              &                         & objects at higher levels \\
\hline level(min(2), max(4))  &                         & Combinations: objects relevant only in a \\
       level(range(2, 4))     & 2 $\leq$ level $\leq$ 4 & certain level range\\
       level(mask(0x1C))      &                         & \\
\hline
\end{tabular}
\label{table:levelid}
\end{center}
\end{table}
\subsection{Levels of Detail}
To specify levels of detail we use the ``level'' tag:
\begin{verbatim}
   Had_Endcap = cylinder(name("Had Endcap"), level(1), ...);
   Had_Endcap += placement(@something, ..., level(max(0)), ...);
   Had_Endcap += placement(@something_else, ..., level(min(1)), ...);
   World += placement(@Had_Endcap, translate(0, 0, 0.5 * (had_b_length - had_thickness)));
   World += placement(@Had_Endcap, rotate(axis("y"), angle(180 degrees)),
                      translate(0, 0, -0.5 * (had_b_length - had_thickness)));
\end{verbatim}
The full level syntax is described in table~\ref{table:levelid}.

\subsection{ID Calculation}
ID calculation (such as CalorimeterID and TrackerID) is generally used for two purposes:
to specify segmentation of a detector---hits with the same ID are combined to a single
energy deposition---and to specify an abbreviated version of the location of an energy
deposition. The problem is: how can one change the amount or method of segmentation
without changing the C++ source code of the simulator? One solution is the ability to specify
the segmentation method in the geometry file and ``interpret'' it inside the simulator.
The GODL API provides a simple, fast, interpreter to do that. In fact, it ``compiles''
the segmentation specification into byte-code, and runs the byte code for each hit
during the simulation. In practice, this is fast enough to not cause a significant
overall performance hit.

As an example let us consider tracker IDs. For the Vertex detector barrel we would use: 
\begin{verbatim}
   Vertex_Barrel.i = cylinder(name("VXD"), ..., 
                              idCode(code(tb_code), data("system", 1), data("id", i - 1)));
\end{verbatim}
The ID calculation in idCode is specified as a string in a ``code'' list object.
The algorithm for the vertex and barrel trackers is:
\begin{verbatim}
   tb_code = "x: fp0 y: fp1 z: fp2 layer: d3 id z H 0x40000000 mul or system 28 bitshift
              or stop"
\end{verbatim}
For the tracker end cap it is:
\begin{verbatim}
   "x: fp0 y: fp1 z: fp2 layer: d3 id 0x80000000 or z H 0x40000000 mul or system 28 bitshift
    or stop";
\end{verbatim}
Here, ``x: fp0'' means that the API routine that evaluates the byte code associated
with the above, expects x to be given in the first floating point ``register''.
Similarly, ``layer'' is provided as an integer in the fourth register.

The code itself is a reverse polish, PostScript-like, language with some limitations
and some extras: some named variables must be provided by the simulator as standard
arguments (x, y, z, layer), and some named variables are provided using ``data'' object
lists in the specification. In the above, H is the Heaviside step function: 1 if the 
argument is positive, 0 otherwise. The language includes a standard set of math functions
that may be used.

There is slightly more work for the calorimeter ID's. For the end caps we have:
\begin{verbatim}
   cal_code_ec = "x x mul y y mul add sqrt z atan2 theta_seg mul _pi div " + standard_code;
\end{verbatim}
For the barrel we have:
\begin{verbatim}
   cal_code_b = "x x mul y y mul add sqrt z atan2 cos 1.0 add theta_seg mul 2.0 div "
                + standard_code;
\end{verbatim}
where standard\_code is: 
\begin{verbatim}
   standard_code = "truncate 11 bitshift y x atan2m phi_seg mul 0.5 mul _pi div truncate
                    or layer 21 bitshift or system 28 bitshift or stop";
\end{verbatim}
Here, atan2m is the same as atan2, except that the result is given in the range 0 to
$2\pi$. We have to add standard argument specifications to this, for example:
\begin{verbatim}
   cal_code_ec = "x: fp0 y: fp1 z: fp2 layer: d3 " + cal_code_ec;   
\end{verbatim}

\subsection{GODL API}
The GODL API consists of four classes: GODLParser, MCode, MStack and MVar.
There are (currently) 11 virtual functions that the API implementer must write.
For example:
\begin{verbatim}
   virtual int constructCylinder(
      const char   *nameForFutureReference,
      const char   *objectName,
      double        innerRadius,// length units: meter
      double        outerRadius,
      double        length,
      const char   *materialRefName,
      const char   *type,
      int           nLayers,
      int           nSlices,
      double        samplingFraction,
      const MStack &IDCode)
\end{verbatim}

Other functions that must be implemented are:
\begin{verbatim}
   constructCone(...),    addField(...),  addPlacement(...),     constructPlacement(...),
   rotate(...),           translate(...), constructElement(...), constructCompound(...),
   constructMixture(...), addMixture(...)
\end{verbatim}
The API reads the GODL file (which typically has extension .godl) and calls the ``construct''
routines to construct objects and placements. It then calls rotate and translate on the
placements, and addMixtures to add materials to the mixtures. Finally it calls
addPlacement to instantiate an actual placement of an object.

\subsection{Status}
The GODL parser and evaluator are essentially complete and the API layer to access the
volume list exists. GODL was first completely implemented in Lelaps V03-23-26,
including levels of detail and hit ID calculation. GODL representations of the
SDMar04 and SDJan03 detectors exist, the latter with two different levels of detail.
As mentioned, a GODL-to-HepRep converter exists and comes with the Lelaps distribution.

\section{FUTURE}
New features that are planned for Lelaps\footnote[1]{Lelaps (storm wind) was a dog with such speed that, once set upon a chase,
he could not fail to catch his prey. Having forged him from bronze, Hephaestus gave him to
Zeus, who in turn gave him to Athena, the goddess of the hunt. Athena gave Lelaps as a wedding
present to Procris, daughter of Thespius, and the new bride of famous hunter Cephalus.
A time came when a fox created havoc for the shepherds in Thebes. The fox had the divine
property that its speed was so great that it could not be caught. Procris sent Lelaps to
catch the fox. But because both were divine creatures, a stalemate ensued, upon which Zeus
turned both into stone. Feeling remorse, Zeus elevated Lelaps to the skies, where he now
shines as the constellation Canis Major, with Sirius as the main star.}/CEPack include
support for combinatorial geometry and the ability for shower continuation into the next volume.
More tuning of hadronic showers is needed. For GODL, planned new features include adding the
remaining standard geometrical shapes and implementing support for combinatorial geometry

\begin{acknowledgments}
This work was supported by Department of Energy contract DE-AC02-76SF00515.
\end{acknowledgments}

\end{document}